\documentclass[review]{elsarticle}
\pdfoutput=1
\usepackage{amsmath, color}
\usepackage{graphicx}

\biboptions{authoryear}

\begin{document}

\begin{frontmatter}

\title{Strain-induced tuning of optical properties of layered Mo$S_2$}

\author[address]{Shubham Bhagat}
\ead{shubh.279@gmail.com}
\author[address]{Shivani Sharma}
\author[address]{Jasvir Singh}
\author[address]{Sandeep Sharma\corref{mycorrespondingauthor}}
\cortext[mycorrespondingauthor]{Corresponding author}
\address[address]{Department of Physics, Guru Nanak Dev University, Amritsar 143005, India}

\begin{abstract}
The sensitive correlation between optical parameters and strain in Mo$S_2$ results in a totally different approach
to tune the optical properties. Usually, an external source of strain is employed to monitor the optical and
vibrational properties of a material. It is always challenging to have a precise control over the strain and
its consequences on material properties. Here, we report the presence of a compressive strain in Mo$S_2$ crystalline
powder and nanosheets obtained via the process of ball-milling and probe sonication. The diffraction peaks in the  X-ray diffraction pattern shift to higher 2$\theta$ value implying a compressive strain that increases with 
the processing time. The absorption spectra, photoluminescence and Raman modes are blue-shifted w.r.t the bulk unprocessed sample. The observed blue-shift is attributed to the presence of compressive strain in the samples. Whereas in thin nano sheets of Mo$S_2$, it is very likely that both quantum confinement as well as strain result in the observed blue-shift. These results indicate that by optimizing the processing conditions and/or time, a strain of desired amount and hence tunable shift in optical properties of material can be achieved.
\end{abstract}

\begin{keyword}
\texttt {Mo$S_2$} \sep Excitonic absorption \sep Photoluminescence \sep 2-D Transition-metal dichalcogenides
\end{keyword}

\end{frontmatter}

\section*{Introduction}
\label{Introduction}
The transition-metal di-chalcogenides (TMDCs) of the type $MX_2$ (where M = Mo, W and X = S, Se etc) are an exciting class of 2-dimensional (2-D) materials which possess intriguing optical, electrical and thermal properties\cite{Beal1976,Wang2012,Mak2016}. 
In bulk form they exhibit an indirect band gap ($\approx 1.2 eV$) and a direct band gap ($\approx$ 1.8 eV) when reduced to a monolayer. Their non-zero direct band gap is larger than silicon (1.12 eV) and considerable
smaller amount of off-state leakage current is expected in devices prepared on TMDCs monolayers. 
Further, the direct band gap values of these materials lies in the visible region of the spectrum and therefore, these 2-D materials are promising for nanoelectronics and optoelectronics devices\cite{Wang2012,Mak2016}. Although, graphene with exceptionally large mobility ($\approx 10^5 cm^2V^{-1}s^{-1}$) \cite{Matt2010,Choi2010}; was the first 2-D material that has been extensively studied, the lack of band gap make it unsuitable for many optical and electronics applications. Due to their ultrathin nature, a most common feature shared by these 2-D materials is that they are quite sensitive to external perturbations such as strain. Whereas the effect of strain on electronic, optical and vibrational properties of graphene is extensively studied \cite{Ni2008,Neto2009,Ming2010,Novo2011}, the theoretically predicted band-gap opening by external strain has still remained elusive \cite{Lee2008,Kim2009}; Therefore limiting the use of graphene in digital devices. On the other hand, 2-D TMDCs like Mo$S_2$ are predicted to have highly strain-tunable optical and electronic properties\cite{Wang2012,Lu2012}. Not only limited to these areas, the strained nanostructures of Mo$S_2$ have also been predicted to possess improved electrical conductivity and a large value of thermopower (250-350 $\mu V/K$) \cite{Bhat2014}.

Thus, the strain has significant influence on electronic, optical as well as thermoelectric properties of the material. Different methods for inducing strain on 2-D materials have been developed. For instance He et.al.
studied the response of Mo$S_2$ under uniaxial tensile strain by using a cantilever device. They observed
a red-shift in band gap and photoluminescence emission spectra of Mo$S_2$ \cite{He2013}. On the other hand, Hui et.al have used an electromechanical device to apply a uniform and controllable compressive strain in trilayer Mo$S_2$ \cite{Hui2013}.
They reported a blue-shift in optical as well as in vibrational spectra of the compressed Mo$S_2$ thin layers.
Thus, a tensile and compressive strain give rise to a red-shift and blue-shift, respectively.

Here, we report a different approach to introduce strain in the samples during the synthesis process.
We prepared samples using ball-milling followed by probe sonication. During the ball-milling zirconium balls apply shear and compression forces on the powdered sample, resulting in exfoliation of multi-layered sample. This process can result in a compressive strain in the layered material. We have observed a negative compressive strain giving rise to a blue-shift in optical properties. These results are in agreement with previously reported experimental and theoretical data.


\section*{Experimental}
\label{Experimental}
Initially, 2 gm of powder was mixed with isopropyl alcohol. Thereafter, the mixture was processed using high energy ball milling for 32 hours. The grinded sample was dried in oven at $80^{\circ}C$. From the dried sample, 1 gm was kept aside and named S1. The remaining sample was mixed with Dimethylformamide (DMF) and processed with a probe sonicator (750 Watt, 20 kHz, PCI-Analytics, India). After probe sonication, suspension obtained was centrifuged at 10,000 rpm for 20 minutes. Thin sheets floating on the surface of suspension were noticed. The sheets were transferred to the glass slide using dip coating method. The remaining sample was dried to obtain the powder. This powder sample so obtained was named S2. The sample prepared by dip coating was named S3. All the structural and optical characterizations were done on the samples S1, S2 and S3.

 For the structural analysis, X-ray diffraction (XRD), Scanning electron microscope (SEM) and high resolution transmission electron microscope (HRTEM) were used. XRD patterns were obtained using Shimadzu 7000 X-Ray Diffractometer (Cu $K_{\alpha}$ radiation). High resolution Transmission electron microscopic images were obtained using JEOL (JEM-2100). 
Optical absorption spectra of the samples was obtained by using Shimadzu UV-Vis 2450 spectrophotometer and photoluminsence emission spectra was recorded using Perkin Elmer LS55 fluorescence Spectrometer. The vibrational modes in synthesized samples  were studied using Renishaw Invia Reflex micro Raman spectrometer using excitation wavelength of 514 nm.


\section*{Results and Discussion}
\label{Results and Discussion}

\subsection*{X-Ray Diffraction}
\begin{figure*}[htp!]
\includegraphics[width=8cm]{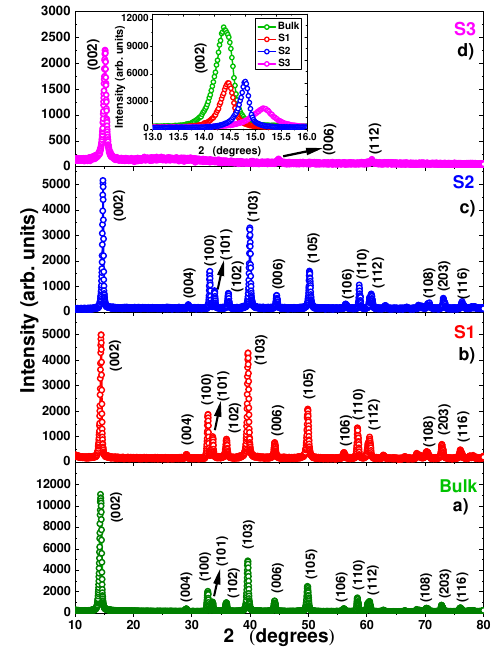}
\caption{(color online)X-ray diffraction pattern from (a)bulk crystalline Mo$S_2$ powder, Sample (b) S1 (c) S2
and (d) S3, thin sheets of $MoS_{2}$}
\label{fig:xrd}
\end{figure*}
Fig.\ref{fig:xrd} compares the diffraction pattern obtained from bulk material, samples S1, S2 and S3. 
The bulk sample shows various diffraction peaks belonging to different planes of highly crystalline Mo$S_2$.
All diffraction peaks can be indexed with crystalline structure (JCPDS 77-1716, P63/
mmc space group, H-Mo$S_2$). It is clear from the data that all samples display intense (002) peak and other peaks of relatively lower intensity. This, confirms the layered structure of the material giving rise to dominant exposure of the (002) planes in all samples. We observe that samples S1, S2 and bulk Mo$S_2$, contain a large number of peaks
in the diffraction spectra, whereas sample S3 consists of three peaks belonging to (002), (006) and (112) planes.
Intense (002) peak indicates the highly layered structure of the deposited nanosheets.  
 
\begin{figure*}
\includegraphics[width=9cm]{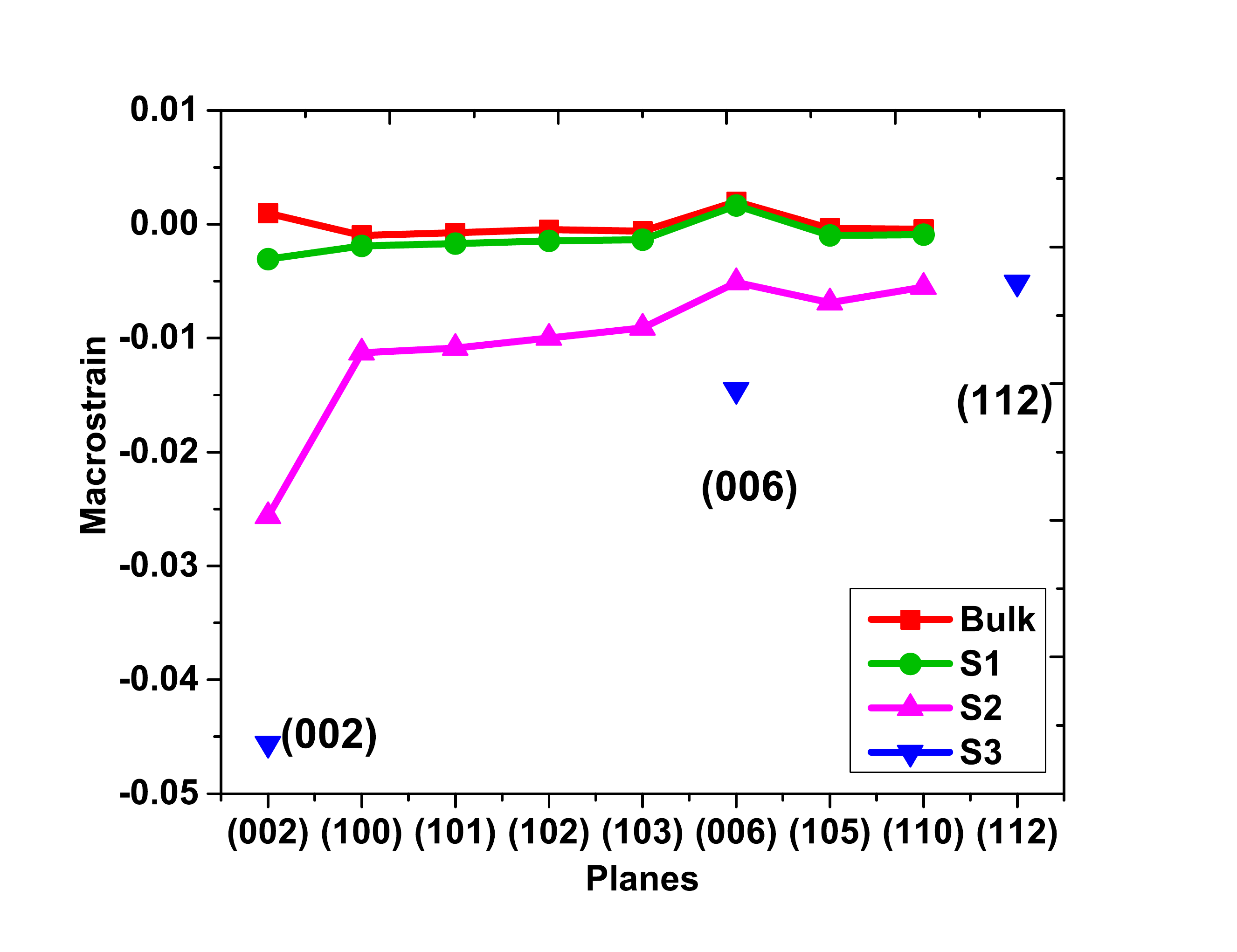}
\caption{(color online)Variation of macrostrain for different samples}
\label{fig:strain}
\end{figure*}

A close analysis of the data reveals [inset in Fig.\ref{fig:xrd}] that all peaks in sample S1 and S2 have shifted towards higher $\theta$ values. This implies that after ball milling and probe-sonication a uniform stress within the sample give rise to a uniform strain. This uniform strain causes a line shift in the peak position. This shift causes a lower d value for the planes. Thus, we can calculate the uniform strain or the macrostrain by comparing the
experimentally determined 'd$_{exp}$' values with the reference values from JCPDS data for the bulk.
The macrostrain is estimated using the relation:
\begin{equation}
macrostrain = \dfrac{d_{exp}-d_{ref}}{d_{ref}}
\end{equation}
where interplanar distance $d_{exp}=\lambda/2\sin\theta$ and $d_{ref}$ is the respective value in JCPDS data\cite{Sinh2012} .
Strain calculations were performed for all peaks and data is shown in Fig.\ref{fig:strain}. It is inferred that bulk crystalline powder display minor positive strain for (002)($0.09 \%$)  plane and negligible negative strain for all other planes. Rest of the samples display a negative compressive strain, which is larger for planes with smaller indices. In sample S3, we observed only three peaks and corresponding compressive strain is larger than other samples. For (002) plane the absolute value is around 4.56$\%$, quite larger from S1 and S2.
Thus, we see that processes of ball-milling and probe sonication give rise to a strain in the sample.

\subsection*{Transmission Electron Microscopy}
\begin{figure*}
\centerline{\includegraphics[width=10cm]{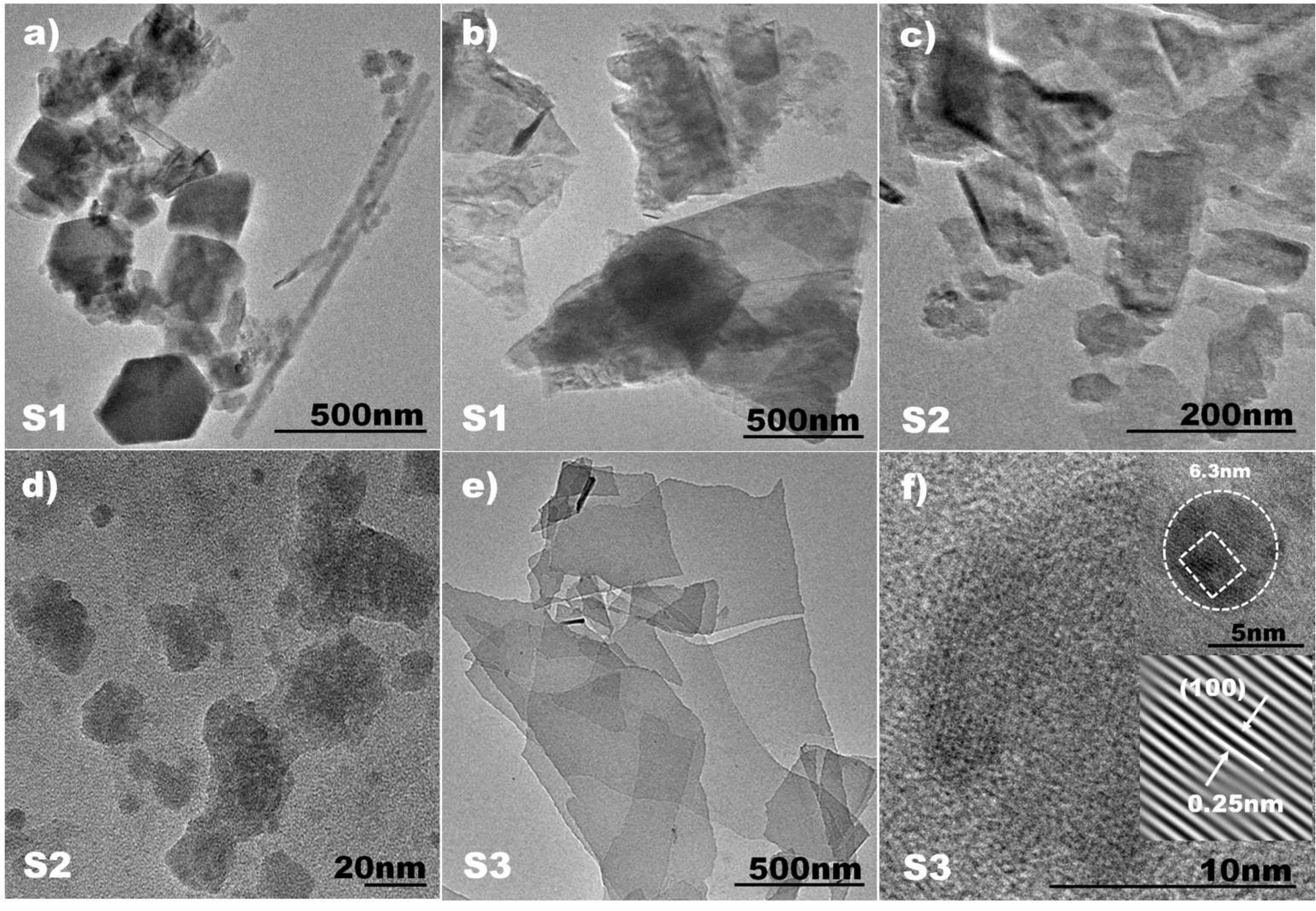}}
\caption{(color online) (a) and (b) Low resolution TEM-images of sample S1, (c) and (d) sample S2,
(e) sample S3 and (f) HR-TEM image of sample S3. Note that in sample S3 ultra thin sheets together with particles of smaller size are obtained}
\label{fig:TEM}
\end{figure*}
\begin{figure*}
\centerline{\includegraphics[width=10cm]{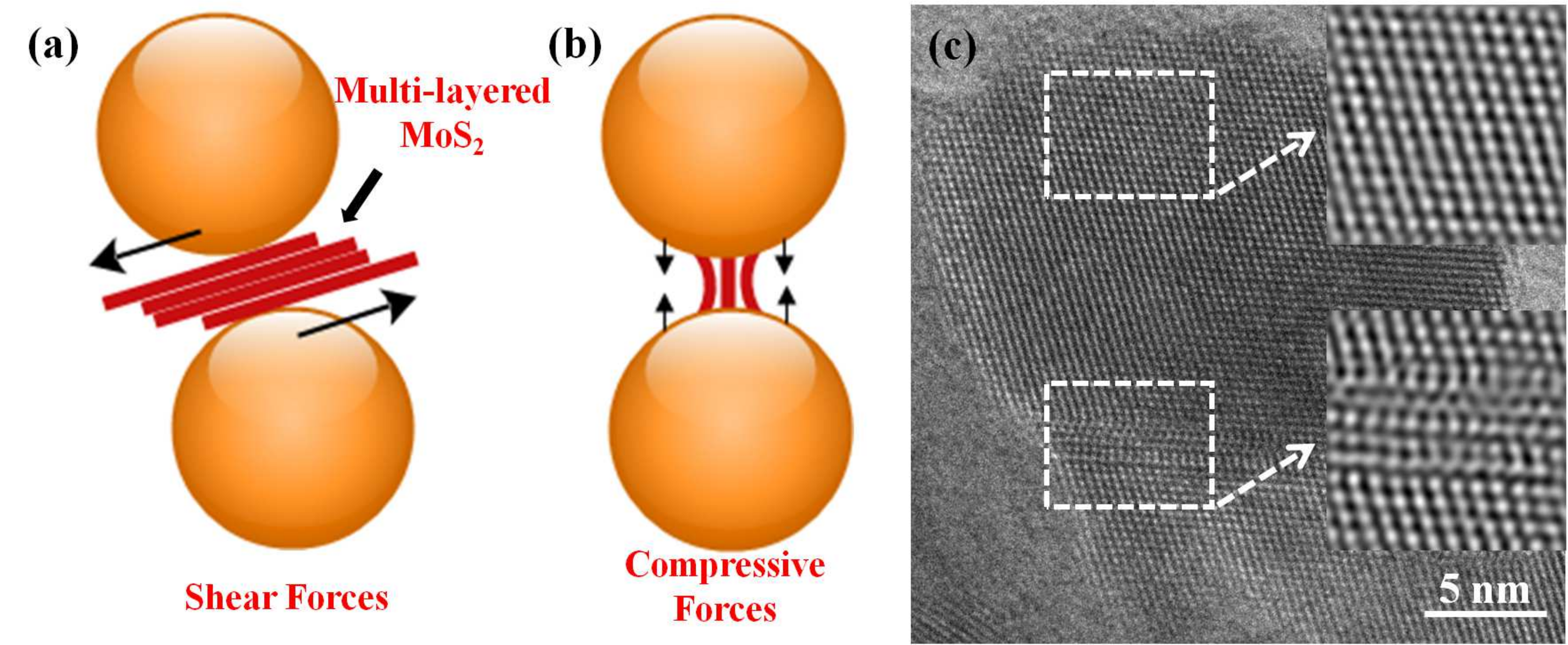}}
\caption{(color online) Fig. (a) and (b) give a representation of shear and compressive forces acting on the sample during ball milling. In (c) HR-TEM image of a thick sheet obtained after ball milling is shown. For more details, refer to the text.}
\label{fig:TEMBallMilled}
\end{figure*}
Fig.\ref{fig:TEM}(a) and (b) show the low resolution transmission electron microscopic (TEM) image of ball milled sample. (c) and (d) are the similar images for sample S2. As we see, the ball-milling has reduced the size significantly from $\mu$ m to a few hundred of nano meters. Particle sizes of various dimensions are obtained.
A significant difference is seen in sample S3, where ultra thin nanosheets together with particles of sub-nanometer sizes are obtained. The upper inset in (f) shows the HR-TEM image of a particle and lower inset displays the fast fourier transformed image of the area shown in upper inset. The lattice fringes with spacing of 0.25 nm corresponding to (100) can be clearly seen.   
The Fig.\ref{fig:TEMBallMilled} (a) and (b) display how the shear and compressive forces reduce the size
of material during ball-milling. More important information is obtained from the HR-TEM image (c) of a thick sheet
of Mo$S_2$. The upper inset in this image represents the digitally filtered image of the highlighted part. The hexagonal symmetry is clearly visible, and the lattice spacing is 0.35 nm. The lower inset represents the FFT image corresponding to lower part of the main image. Interesting features are visible from this image. As we see, the continuous
running line are disrupted. This is indeed due to the compressive forces acting on the layered sample (Fig.\ref{fig:TEMBallMilled} (b)). This results in a compressive strain in the sample and as a result,
the spacing between two adjacent (compressed) planes is $\approx$ 0.33 nm. These results corroborate the 
XRD data where, we have noticed that diffraction peaks have shifted towards higher $\theta$ values, thus
suggesting the presence of a compressive strain in the samples. 

\subsection*{Raman Spectroscopy}
\begin{figure*}
\includegraphics[width=10cm]{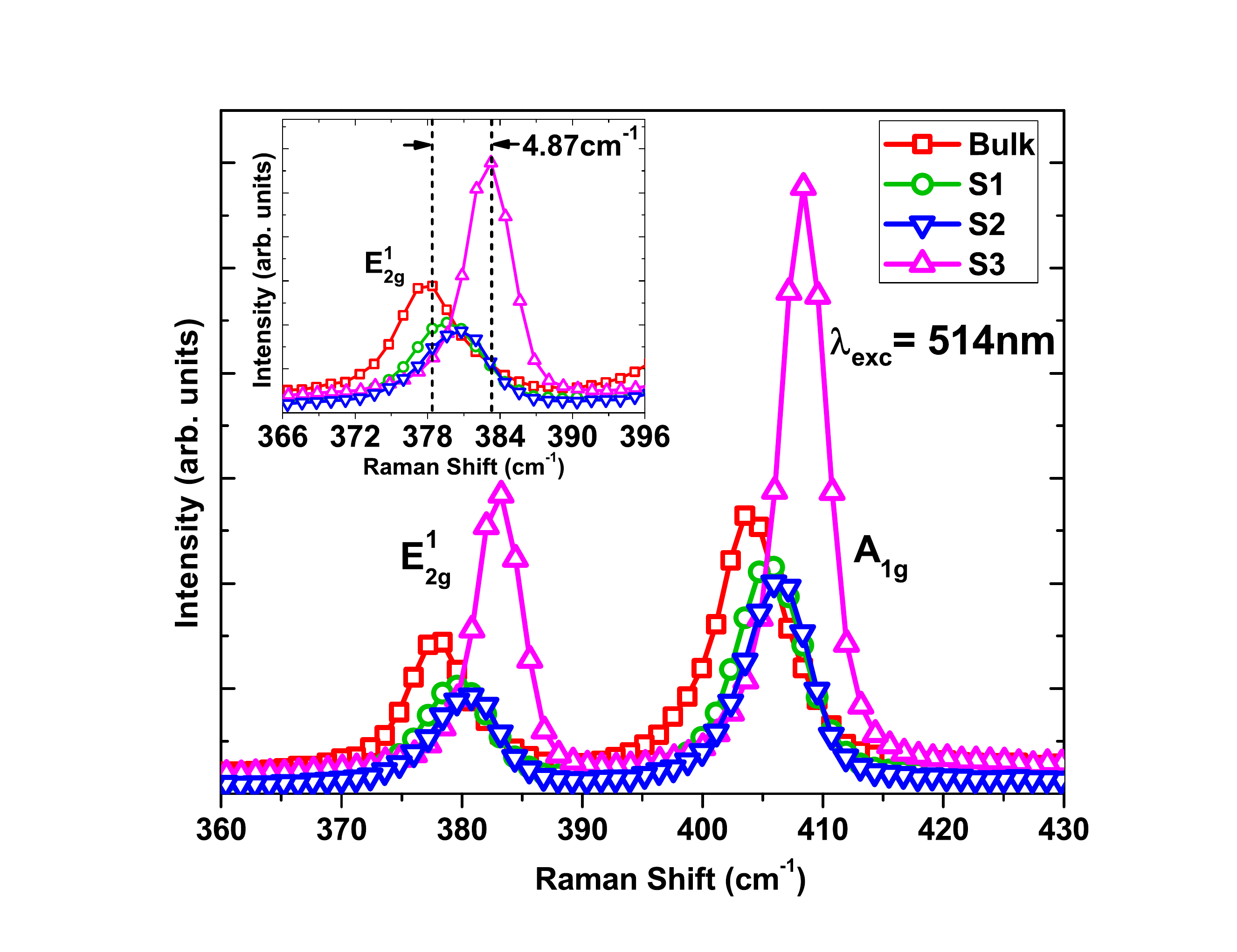}
\caption{(color online)Comparison between raman spectra obtained from bulk crysrtalline powder, ball milled powder and thin sheets  at 514 nm excitaion wavelength. A blue-shift is clearly visible in the two modes.}
\label{fig:raman}
\end{figure*}
Fig.\ref{fig:raman} displays the typical Raman spectra of bulk, ball milled, probe sonicated sample and Mo$S_2$ nanosheets.  The spectra depicts the only two characteristic Raman active modes: $A_{1g}$ and $E^1_{2g}$ which were obtained with excitation wavelength of 514 nm. Here, $A_{1g}$ is the mode that arise due to out-of-plane motion
of the Mo and S atoms. The $E^1_{2g}$ mode arises due to the in-plane motion of the Mo and S atoms \cite{Wang2012,Shar2017} .
In bulk sample, the in-plane mode $E^1_{2g}$ appears at 378.39 $cm^{-1}$, whereas the out-of-plane mode at 403.52 $cm^{-1}$. The separation between the two modes is 25.13$cm^{-1}$, in good agreement with that found in mechanically exfoliated multilayered Mo$S_2$\cite{Plec2012}. Both of these modes have shifted to higher frequencies
in processed samples. In sample S1, the $A_{1g}$ and $E^1_{2g}$ mode have shifted by 2.39 $cm^{-1}$ and 1.19 $cm^{-1}$, respectively, with a shift ratio of $A^{1g}$/$E^1_{2g}$ of $\approx$ 1. As we notice in the 
Table.\ref{table_514nm}, the separation between these characteristics modes is almost similar in all samples.
But, the sample S3 with nanosheets, has largest displacement w.r.t. the bulk sample. For e.g., the modes $A_{1g}$
and $E^1_{2g}$ have shifted by 4.85$cm^{-1}$ and 4.87$cm^{-1}$, respectively, with a shift ratio of $\approx$ 1. Therefore, we notice that in ball-milled and probe sonicated samples the modes undergo a blue-shift. The sample
with largest compressive strain shows largest blue-shift. Such a blue-shift in these two modes has previously been noticed in Mo$S_2$ bulk crystals and nanotubes under high pressure and was attributed to the presence of compressive strain \cite{Hui2013,Bagn1980,Virs2007}.  

\begin{table}
\caption{Peak positions for characteristic Raman modes A$_{1g}$ and E$^1_{2g}$ of MoS$_2$ in different samples at an excitation wavelength of 514nm. The last column gives the separation between these two modes.} 
\label{table_514nm} \vspace*{4mm}
\begin{tabular}{ c  c  c   c  }
\hline \hline
Sample   & $A_{1g} \left( cm^{-1}\right) $  & $E^1_{2g} \left( cm^{-1}\right) $  & $A_{1g}-E^1_{2g} \left( cm^{-1}\right) $ \\
             \cline{2-4}
\hline
\hline
Bulk     & 403.52            & 378.39              & 25.13 \\
Ball-milled  & 405.91        & 379.58              & 26.33 \\
Sonicated & 405.92     & 380.8              & 25.12 \\
Nanosheets     & 408.37      & 383.26              & 25.11 \\
\hline
\hline
\end{tabular}
\end{table}

\subsection*{UV-Vis Spectroscopy}
\begin{figure*}[htp!]
\centerline{\includegraphics[width=8cm]{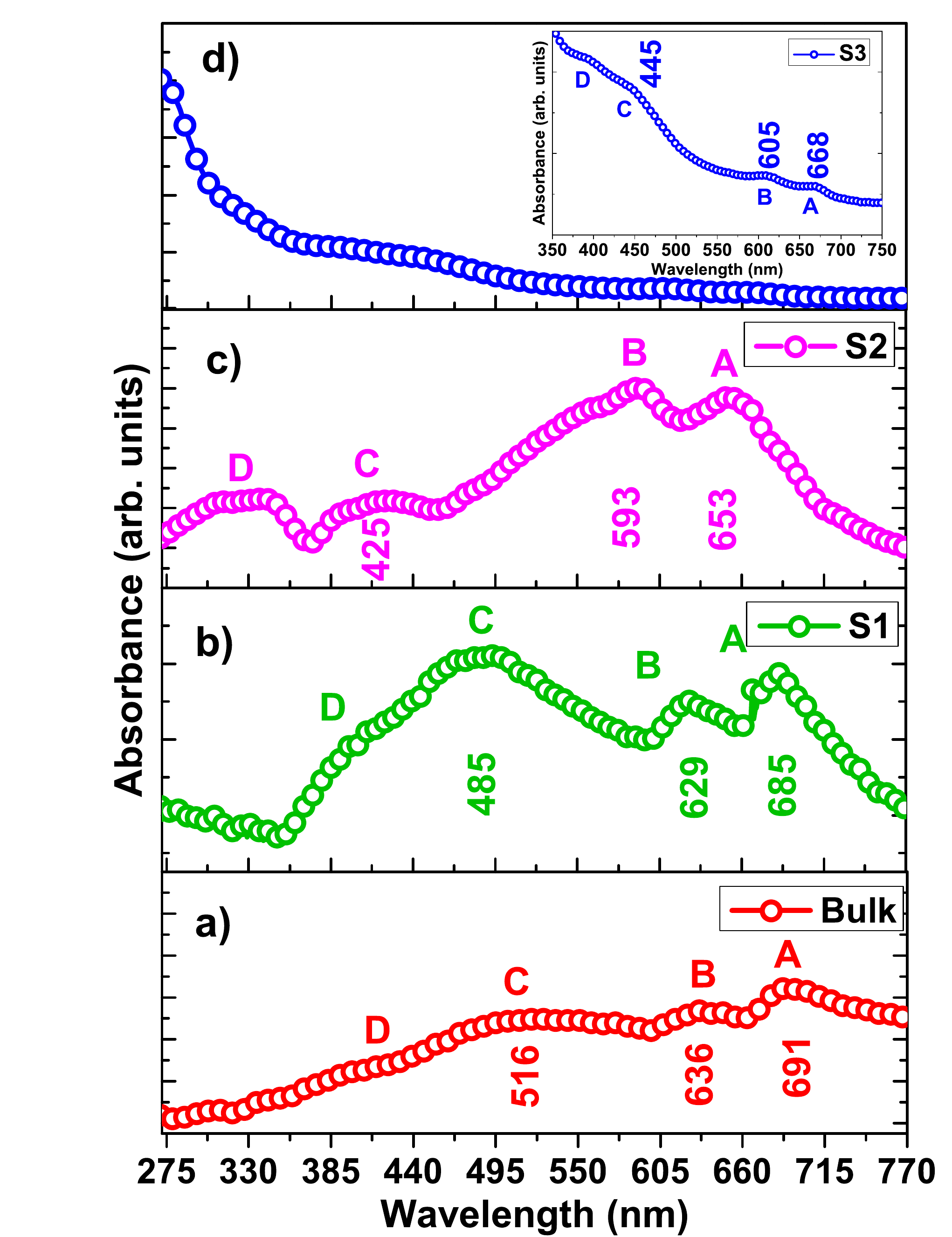}}
\caption{(color online) Absorption spectra of bulk $MoS_{2}$ and samples S1, S2 and S3.}
\label{fig:uv}
\end{figure*}
Fig.\ref{fig:uv} compares the absorption spectra obtained from bulk crystalline Mo$S_2$, ball milled sample S1 and probe sonicated sample S2. The absorption spectra from nanosheets (S3) is also shown for comparison. The absorption spectra of the powder samples were obtained by using the diffuse reflectance geometry. Initially, the absorption coefficient was calculated using the Kulbeka-Munk equation;
\begin{equation}
F(R) = \frac{(1-R)^2}{R} = \frac{k}{s}
\end{equation} 
where, F(R) is the Kulbeka-Munk function (which is the required absorption coefficient $\alpha$), R is the absolute reflectance of sample, k is the molar absorption coefficient and s is the scattering coefficient. The plot of F(R) against wavelength gives the required absorption spectra \cite{Murp2007,Wood1979,Rose2012}.

The absorption spectra for the bulk powder is shown in Fig\ref{fig:uv}(a). The peaks marked as A (691 nm) and B (636 nm), are the characteristic excitonic peaks of Mo$S_2$. These peaks originate from transition between conduction band and spin-orbit-coupling (SOC) induced splitting of valence band energy levels. These peaks are separated by $\approx$ 156 meV, quiet close to the theoretically calculated value (160 meV) \cite{Mukh2015}. However, it should be note that there exist a wider spread in the reported values of spin-orbit coupling strength in Mo$S_2$ \cite{Brom1972,Boke2001,Zhu2011,Dani2017}. For a single layer Mo$S_2$, splitting can entirely attributed to the SOC. However, in bulk it might be due to the combination of spin-orbit coupling and inter-layer coupling. But in literature, there is a disagreement between the relative strength of these two mechanism \cite{Klei2001,Jin2013,Moli2013,Nass2014,Suzu2014,Ekna2014,Padi2014,Latz2015}.

\begin{table}
\caption{Summary of various peak positions in absorption spectra of different samples. Spin-orbit-coupling induced
splitting between peak position A and B is also given for the comparison} \label{table_UV_vis} \vspace*{4mm}
\begin{tabular}{ c  c  c   c  }
\hline \hline
Sample   & A (nm)  & B (nm)  & A-B (meV) \\
             \cline{2-4}
\hline
\hline
Bulk     		& 691 & 636  & 156 \\
Ball-milled  	& 685 & 629  & 162 \\
Sonicated 		& 653 & 593  & 190 \\
Nanosheets     	& 668 & 605  & 189 \\
\hline
\hline
\end{tabular}
\end{table}
 
The other peaks marked as C and D are attributed to the direct transition from the depth of the valence band \cite{Wilco1995, Wils1969}. The absorption spectrum from samples S1, S2 and S3 also display multiple spectral features that are blue-shifted w.r.t. the bulk. Therefore, after processing the different Mo$S_2$ samples
preserve the excitonic features and are blue-shifted. The maximum energy shift is close to $\approx$ 30 meV, comparatively much smaller than the value ($\approx$ 780 meV) previously reported for Mo$S_2$ nano clusters with sizes in the range of 4.5 nm \cite{Wilco1995, Wilco1997}. This was attributed to the phenomena of quantum confinement in the Mo$S_2$ nanostructures \cite{Wilco1995,Gan2015}. Another noticeable difference in sample S3 appears at lower wavelengths. Below 330 nm, the absorption rises and is attributed to the band edge absorption from scaled nanosheets of Mo$S_2$. It should be noted that quantum confinement effects come into picture when the particle size is close to the Bohr radius, 'R' which is different for different materials.
The Bohr radius is expressed as 
\begin{equation}
R = \epsilon \left( \dfrac{m_o}{\mu} \right) a_0
\end{equation} 
where, $\epsilon$ is the dielectric constant of material in bulk form, $m_0$ is the free electron mass, reduced mass of the exciton $\mu = \dfrac{m_e.m_h}{m_e + m_h}$, and $a_0$=0.53\AA. 
For Mo$S_2$, the effective masses for electron and holes are; $m_e=0.48m_0$ , $m_h= 0.41m_0$ and $\epsilon$=11 \cite{Mukh2015} and this gives $\mu \approx 0.22m_0 $. As a result the exciton Bohr radius for Mo$S_2$ is around 2.65 nm. Thus, when particle size is close to the Bohr radius, one can expect the enhancement in the band gap energy\cite{Gopa2014} as well as increase in the excitonic absorption energy\cite{Wilco1995}, thus giving a blue-shift in absorption energies. The indirect band gap for bulk Mo$S_2$ is close to 1.23 eV and it changes to 1.89 eV when in monolayer form. Therefore, in sample S3, the band edge absorption close to 330 nm (3.8 eV) cannot be explained without invoking the quantum confinement phenomena in the nanosheets. Whereas, in samples S1 and S2 the particle size is quite far away from the quantum confinement regime, the observed blue-shift is indeed not due to confinement phenomena in the nanostructure. But, it is the presence of the compressive strain in these sample, which governs this blue-shift. Recently, He et.al. \cite{He2013} has shown that applying a uniaxial tensile strain to atomically thin Mo$S_2$, causes a red-shift $\approx$ 70 meV for direct gap transition. On the other hand, Hui et.al, has reported the blue-shift of the direct band gap, photoluminescence and Raman active modes under the influence of external compressive strain. Similar, theoretical arguments have appeared justifying the role of strain in controlling the optical and electrical response of the Mo$S_2$ \cite{Hui2013,Swas2014,Weit2015,Keli2013}.

 Hence, the observed blue-shift in samples S1 and S2 is due to strain in the samples whereas in sample S3 it is under the combined effect of compressive strain and quantum confinement effects. 

\subsection*{Photoluminescence Spectroscopy}
\begin{figure*}
\centerline{\includegraphics[width=8cm]{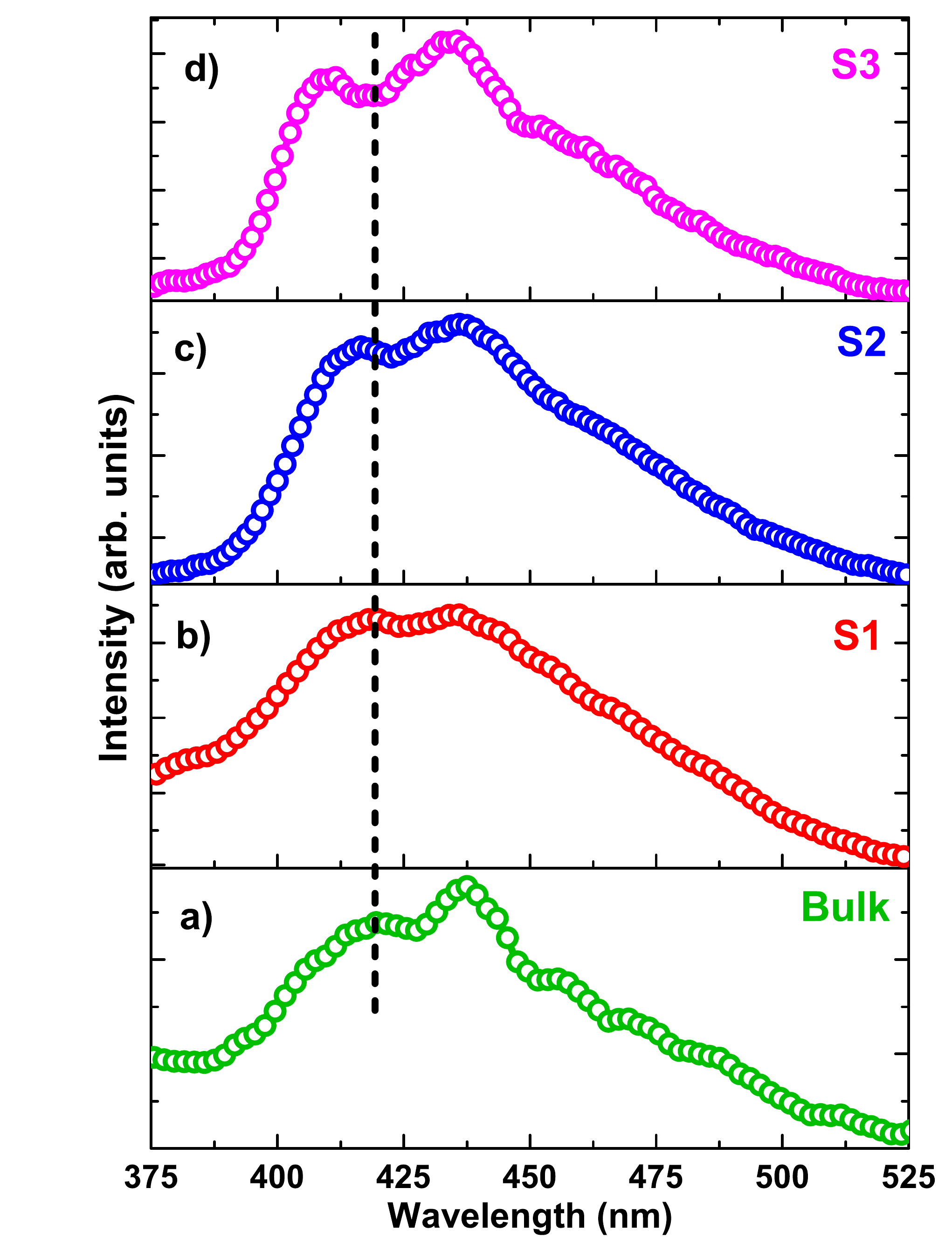}}
\caption{(color online)Photoluminescence spectra from $MoS_{2}$ nanostructures at 320 nm excitaion wavelength for (a) Bulk crystalline powder (b) Ball Milled powder (c) probe sonicated powder and (d) probe sonicated suspension}
\label{fig:pl}
\end{figure*}

Fig.\ref{fig:pl} displays the photoluminescence spectra of different samples acquired using excitation wavelength of  320 nm. Bulk sample, displays a relatively weak and broad emission covering the range of C and D-type absorption.
Other samples also display similar spectral features except that they are blue-shifted w.r.t. the bulk spectra.
Therefore, as the strain has increased, the emission peaks in the PL spectra shifted to higher energy.


\section*{Conclusion}
Different samples of Mo$S_2$ were prepared using high energy ball-milling followed by probe-sonication in DMF. The diffraction peaks corresponding to various planes were shifted towards higher 2$\theta$ values, implying the presence of compressive strain in ball-milled and sonicated samples. The maximum strain is observed in sample S3, which was obtained using
ball-milling followed by probe sonication. The absorption spectra indicated the presence of various excitonic
features, that are blue-shifted w.r.t the bulk sample.  This is further corroborated by the blue-shift
in the PL emission and in the characteristic, $E^1_{2g}$ and $A_{1g}$ Raman modes. The observed blue-shift in powdered samples cannot be associated with quantum confinement effect as their size is quite away from the confinement regime. Therefore, we conclude that the compressive strain is responsible for the observed blue-shift in the absorption/emission and Raman modes. However, the possibility of quantum confinement cannot be excluded in Mo$S_2$ nanosheets, where the sheet size in nano regime is also obtained. Thus, we conclude that strain engineering offers a new route to tune the structural as well as optical properties of material. And this is not only limited to optical but other physical properties like magnetism \cite{Zhou2012,Pan2012} can also be influenced by tunable strain.   


\section*{Acknowledgment}
One of the authors Shubham Bhagat acknowledges the GNDU-Amritsar, India for providing the fellowship under UPE-Scheme for carrying out her PhD work. The authors gratefully acknowledge UGC- New Delhi for providing financial assistance under the project No.F.30-137/2015(BSR).  

\section*{References}
\bibliography{Paperref_used_in_paper}
\end{document}